# A Low-Frequency Vibration Experimental Platform for University Physics Experiment Designed by LabVIEW

Yangjie Dai , Leijian Wang, Wenbin Wu, Aiping Chen and Dawei Gu*

School of Physical And Mathematical Sciences, Nanjing Tech University, Nanjing, China
E-mail: yangjiedai@njtech.edu.cn, ljwang@ihep.ac.cn, 202021130024@njtech.edu.cn, chenaiping@njtech.edu.cn and dwgu@njtech.edu.cn

**Abstract:** Virtual instrument technology has been increasingly used in university physics experiment teaching. An experimental platform is specifically constructed for studying low-frequency vibrations in university physics, which is based on a computer and its internal sound card, along with a program developed in LabVIEW programming environment to perform control and measurement on our experimental platform. The proposed platform effectively replaces the conventional signal generator and oscilloscope traditionally used in such experiments by integrating virtual instruments and essential experimental equipment. The platform offers various functionalities, such as synchronous transmission and reception of low-frequency signals, frequency measurement, dynamic frequency sweep measurement, and measurement using the three-point approximation method. The proposed platform has been successfully applied in experiments involving forced vibration, resonance of tuning forks, and dynamic measurement of Young's modulus. Unlike conventional low-frequency vibration experiments, the proposed experimental platform optimizes efficiency, reduces costs, and offers opportunities for enhancing the instructional content of experiments. Furthermore, the incorporation of state-of-the-art computer technology enhances students' engagement and enthusiasm for learning.

## 1. Introduction

As versatile electronic instruments widely used in scientific research and engineering design, oscilloscopes and signal generators have found extensive applications in university laboratories for science and technology [1-3]. These instruments are commonly utilized in university physics labs to perform experiments involving dynamic measurement of Young's modulus [4], investigation of vibration and forced oscillation [5], and exploration of string vibrations [6]. In these experiments, the signal generator acts as the driving source, inducing periodic mechanical vibrations in the experimental samples. The sensors then convert the mechanical vibrations into electrical signals through specific physical changes, which are subsequently observed using the oscilloscope. Studying experimental content and principles entails comparing variations in signal amplitude and phase displayed on the oscilloscope. Conventional teaching approaches for these experiments traditionally rely on students' observations and manual recording of experimental data to interpret the experimental results displayed on the oscilloscope. However, such an approach may be susceptible to students' subjective judgments, potentially affecting the interpretation of objective experimental facts. Moreover, the lack of effective and practical data recording methods in the classroom hinders the timely preservation of a substantial amount of generated experimental data, thereby impeding further analysis and in-depth exploration of the experiments. Moreover, this issue impacts students' comprehension and mastery of the fundamental concepts related to these experiments.

With the rapid advancement of technology, the domain of fundamental experimental instruction has progressively adopted digital instruments and devices. These tools offer an array of communication interfaces, enabling computerized control and data acquisition. In the context of university physics laboratory instruction, numerous academic institutions have also begun to explore the integration of computers. By employing programmatically controlled instruments, they seek to enhance the pre-existing experimental techniques, foster innovative thinking skills, and enhance the breadth and depth of physics experimental teaching. The goal of such integration of technology is to amplify students' sense of curiosity, ignite their thirst for knowledge, and foster their capacity for innovation.

Fernando Reyes-Aviles utilized augmented reality technology along with mobile devices, measurement instruments, and a breadboard circuit to develop an experimental system for understanding resistive circuits and enhancing students' comprehension [7]. L de la Torre employed internet connectivity and Java-compatible browsers as a means of remotely conducting experiments in university physics education. The focus was on topics such as spring elasticity and the principles of reflection and refraction [8]. P. J. Moriarty designed and implemented instructional courses utilizing the application designed by LabVIEW to explore signal processing, waveform analysis, electrical phenomena, and mechanical resonance [9]. Jinda Sinlapanuntakul developed a software using LabVIEW for controlling the sound card in personal computers, creating virtual oscilloscopes and signal generators. These tools were effectively utilized in RLC circuit experiments, leading to a substantial reduction in experimental expenses [10]. Huang Ningman designed a data processing program in C language specifically for viscosity measurements using the drop ball method [11]. This innovative program played a pivotal role in enhancing experimental efficiency, streamlining the data analysis process, and ensuring accurate results in viscosity measurements. Lan Yanna utilized Visual Basic programming to conduct data processing and error analysis for rotational inertia experiments using the torsion pendulum method. This approach led to enhanced accuracy in the processing of experimental data [12]. Liang Limin developed a Hall effect testing system using the software designed by LabVIEW, seamlessly integrating Keithley instruments and electromagnetic hardware. The primary goal of this system was to address the limitations commonly encounter with conventional Hall effect testers [13]. Wang Jianzhong enhanced the experimental verification of Mariotte's law utilizing the PASCO system. This improvement involved the utilization of the LabVIEW program-controlled Science Workshop 750 data acquisition system and the concurrent validation of Mariotte's law through the employment of the CI-6504 optical sensor [14]. Weikai Ren, from the Institute of Energy Research at Peking University, introduced a PSGCN network, which is grounded in GCN, for the analysis of chaotic time series and prediction of flow parameters in gas-liquid two-phase flows. He further designed a complex vision-family graph network utilizing the phase space visibility algorithm, and proposed a novel multiphase flow measurement tool equipped with a digital ICS sensor. [15-17]

The above-mentioned improvements primarily revolve around the seamless integration of digital instrumentation with computer systems and the development of experimental programs. These advancements have facilitated numerous benefits including real-time display, data acquisition, storage, and automated signal measurements. A considerable portion of these experimental programs, explicitly tailored for experimental setups, is developed using the renowned LabVIEW graphical programming language provided by National Instruments (NI) Corporation. This serves as a compelling testament to NI's dedication to advancing virtual instrumentation technology's application in the realm of scientific research and experimentation. Virtual instrumentation represents a harmonious amalgamation of computers, instruments, measurement and control systems and testing software. Through the utilization of various software programming techniques, virtual instrumentation technology seamlessly integrates multiple hardware components to achieve the realization of numerous instrument functionalities. Thus, it encompasses a wide range of experimental requirements, catering to a broad spectrum of scientific and research needs. Embracing virtual instrumentation technology as the prospective path for instrument development not only results in cost savings but also enhances the standardization and modularity of experimental instruments. This technology surpasses the constraints of conventional instruments, frequently marked by singular functionality and rigid configurations. Consequently, this technology paves a promising path towards the establishment of exceptionally advanced physics laboratories in esteemed universities.

In university physics experiments focusing on low-frequency vibrations, the typical experimental setup involves essential components such as a signal generator, oscilloscope, test rig, and specimen. To enhance these instruments using the capabilities of virtual instrumentation, computer programs need to be utilized for controlling the signal generator and oscilloscope during the experiment. The widespread adoption of multimedia technology has turned computer sound cards into everyday devices that excel as audio signal acquisition systems. The digital signal processors of these cards incorporate analog-to-digital converters (ADCs) and digital-to-analog converters (DACs). The ADCs are responsible for capturing audio signals, while DACs are utilized to reproduce these digital sounds. Numerous virtual instruments have been developed by leveraging computers and sound cards [10, 18-20]. Therefore, this study employs a program designed by LabVIEW in conjunction with the hardware capabilities of the computer, specifically the sound card，to develop an advanced virtual instrumentation platform for conducting physics experiments in university laboratories without relying on conventional signal generators and oscilloscopes. The proposed platform facilitates experiments covering measurements of dynamic modulus of elasticity, forced vibrations, resonance, and other low-frequency signal-related physics projects. This platform enables real-time display, acquisition, analysis, and storage of measured signals. The application of this platform in university physics laboratory teaching frees students from the burden of spending substantial time on manual data recording. Instead, it empowers them to concentrate on data analysis, exploration of experimental patterns, and attaining comprehensive insights into the experimental content.

## 2. Experimental Platform Hardware Design

Figure 1 illustrates the framework of our low-frequency vibration test platform. The red dotted line highlights the core component of the platform, which consists of a platform program developed by LabVIEW. This program controls the computer sound card, enabling signal generation and acquisition without the need for signal generators and oscilloscopes. Through the integration of software and hardware, an efficient university-level platform is designed specifically for conducting low-frequency vibration experiments in the field of physics. The proposed platform is composed of a conventional desktop computer running on the Windows 11 operating system. The computer motherboard is equipped with the Realtek ALC662 audio chip, offering support for 16/20/24-bit SPDIF output and a maximum sampling rate of 96 kHz. The sound card features the TDA2030A module, which serves as the audio amplifier. With an operating voltage range of 6-12 V, the TDA2030A module effectively amplifies low-power audio signals, achieving a single-channel power amplification of 18 W. The mechanical components of the experimental platform consist of the FB823 Forced Vibration and Resonance Experiment Apparatus (manufactured by Hangzhou Jingke Instrument Co., Ltd.) and the Young's Modulus Measurement Instrument (manufactured by Nanjing Langboke Educational Instrument Co., Ltd.).

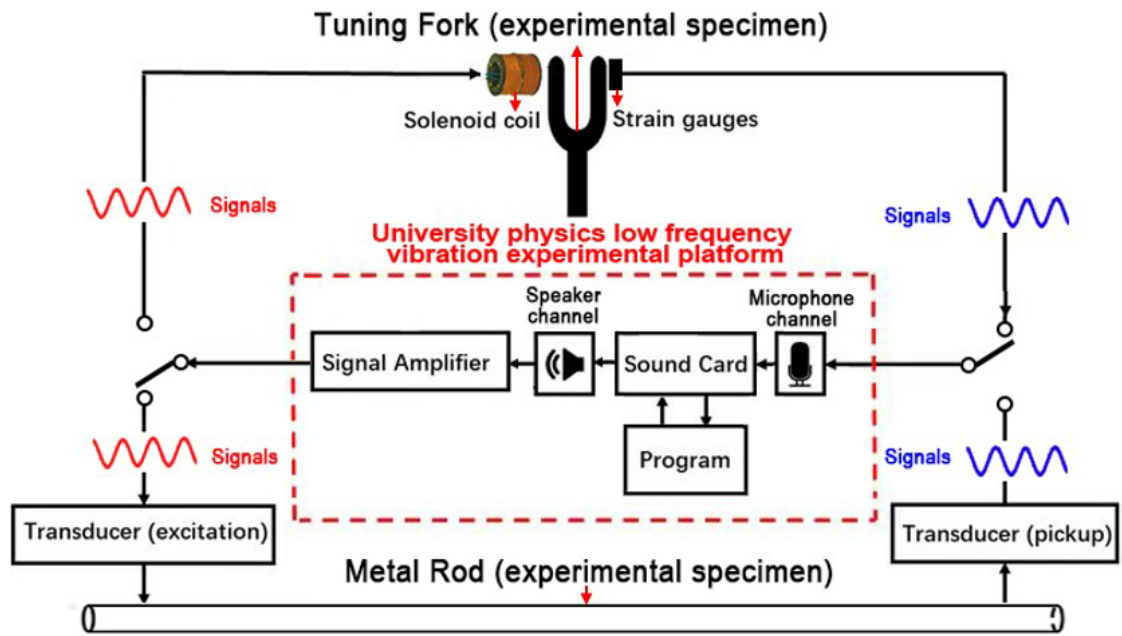

**Figure 1.** Overall hardware structure of low-frequency vibration experimental platform for university physics experiment.

## 3. Experimental Platform Program Design
### 3.1. Feasibility Analysis of Sound Card Signal Transmission and Reception

In this study, a series of experimental tests were perfomed to examine the utilization of sound card in low-frequency vibration experiments and assessed their suitability for university physics experiments that involve low-frequency vibrations. The comprehensive experimental results indicate that the signal acquisition performance of the sound card meets all the requirements for conducting low-frequency physics experiments within the frequency range spanning from 0 to 1200 Hz. For a more detailed overview of the experiment, please consult Supplementary 1.

### 3.2. Overall Program Design

In traditional low-frequency vibration experiments, the excitation signal frequency is typically manually tuned using a signal generator, and the resonance frequency is determined by observing the relationship between the amplitude of the pickup signal and the frequency through an oscilloscope. However, the traditional manual measurement method becomes time-consuming with the increase in the scope of experiments, limiting the ability to conduct more advanced experiments. To enhance experimental efficiency, the experimental platform designed in this study incorporates a dynamic sweep frequency measurement mode alongside the established fixed frequency measurement function. Figure 2(a) illustrates the flowchart of the experimental platform program for university physics low-frequency vibration experiments.

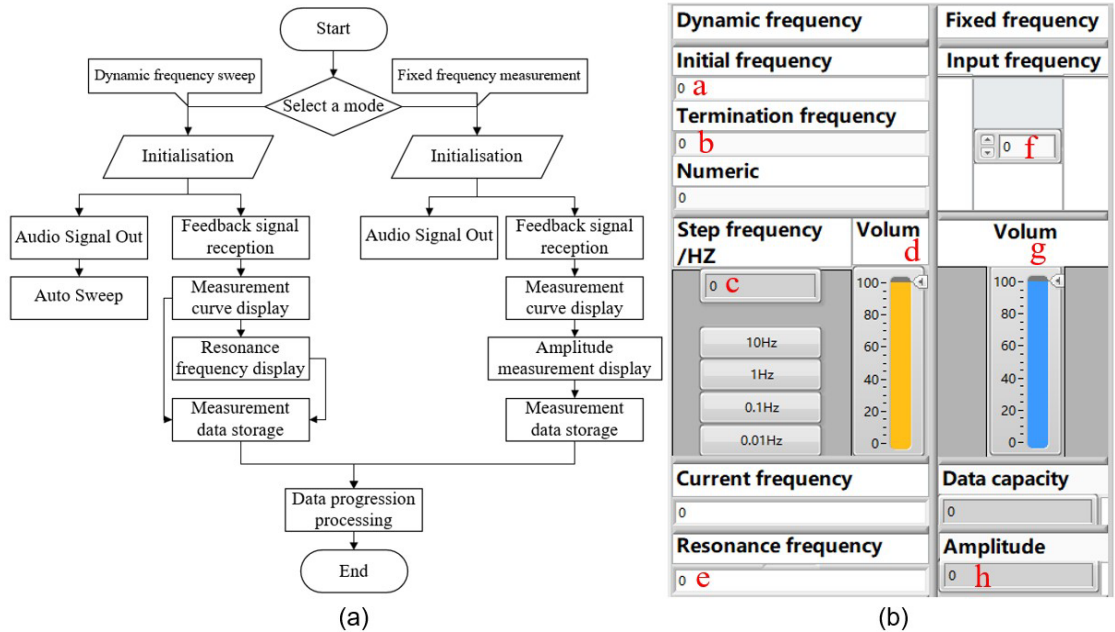

**Figure 2.** Overall program design of low-frequency vibration experimental platform for university physics experiment (a) Flow chart of program operation (b) Image of user interaction front panel.

The user interaction front panel is depicted in Figure 2(b). In the dynamic sweep frequency mode, the users can customize parameters based on their preferences, including the initial frequency (a), final frequency (b), frequency change step (c), and volume level (d). The platform incorporates the user's selected frequency range settings and accurately commands the sound card to generate sweep signals at specific frequency intervals for the experimental specimen. The sound card receives and processes the pickup signal data, leading to the acquisition of a curve representing the relationship between the amplitude and frequency of the specimen's vibration. Consequently, the resonance frequency (e) is identified. On the contrary, in the fixed frequency measurement mode, users only need to input two parameters: the output frequency (f) and the volume level (g). Subsequently, the program controls the sound card to emit the desired fixed-frequency signal to the experimental specimen. Once the pickup signal stabilizes, the corresponding amplitude (h) can be obtained.

### 3.3. Synchronization of Signal Transmission and Acquisition

In university physics experiments, signal generation and acquisition are typically accomplished with two separate instruments: a signal generator and an oscilloscope, respectively. However, in the developed experimental platform, the sound card's playback and recording channels collaborate harmoniously to handle both signal generation and acquisition. Consequently, when designing the experimental program, it becomes crucial to address the synchronization of the sound card's output and acquisition signals. In the LabVIEW development environment, NI Corporation offers various mechanisms for data synchronization, including event occurrence, collection points, and queues. In this study, the synchronization between the sound card's output and acquisition signals was ensured by employing collection point technology [21].

The "Wait at Rendezvous" function plays a crucial role in the data synchronization mechanism of the collection point. This function enables the thread to pause until the predetermined number of tasks specified at the collection point is reached, ensuring the continued flow of data. The program block diagram in Figure 3(a) illustrates the collection point data synchronization mechanism used in this study for the synchronization between the sound card's output and acquisition signals. The program establishes two tasks: signal generation and signal acquisition. Both "Wait at Rendezvous" functions must receive the "Handle Data Stream" signal or the concurrent execution of signal generation and signal acquisition tasks by these two functions.

An audio cable was employed to establish a direct connection between the sound card's playback and recording channels. Direct signal capture from the playback channel was achieved by controlling the recording channel. Consequently, the synchronization effect between the sound card's output signal and the captured signal within the experimental program was evaluated. Figure 3(b) compares the waveforms of the transmitted and received signals of the sound card. The graph in the figure reveals a phase difference of 0.22 ms between the output and captured signals. Throughout the experimental process, it was determined that the phase difference, which had no impact on the obtained results. These observations confirm that the synchronization mechanism implemented in the LabVIEW-based experimental program fulfills the demands of the experiments.

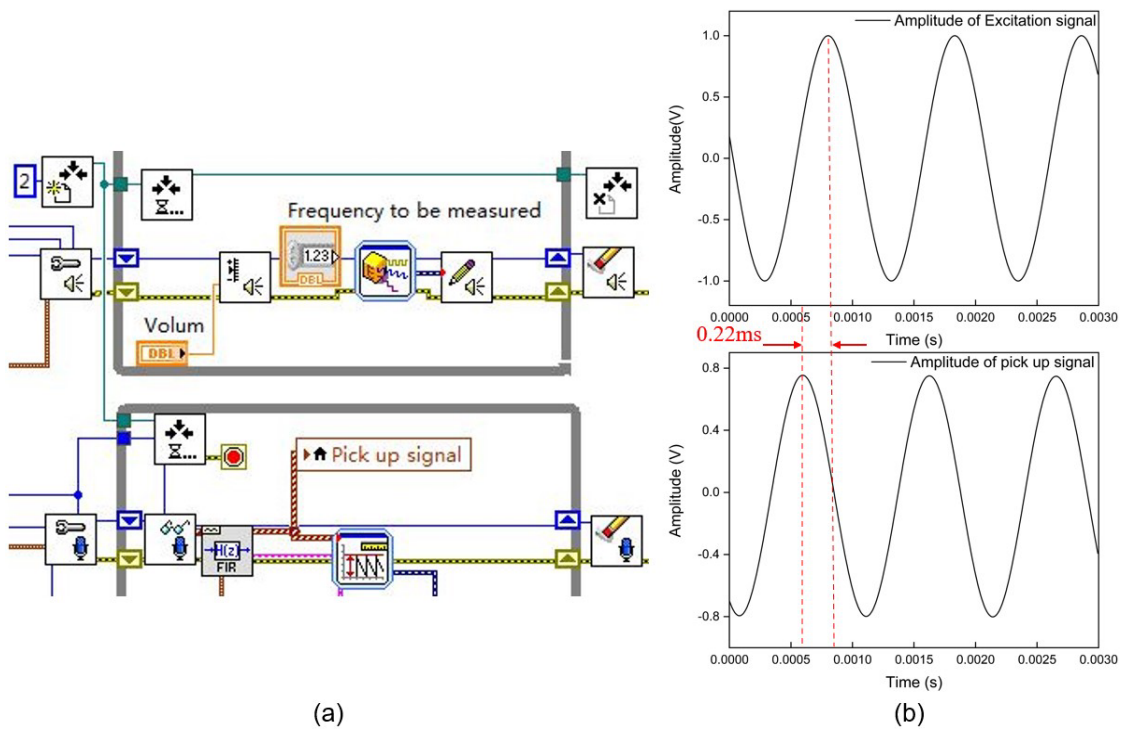

(a)           (b)

**Figure 3**. Signal transceiver synchronization test of the sound card (a) Block diagram of the synchronization mechanism program (b) Comparison of the waveforms of the transmitted and received signals of the sound card.

## 3.4. Determination of Received Signal Amplitude

In the field of vibration signal analysis, frequency domain analysis is frequently employed to identify resonance frequencies. However, this method exhibits a relatively large error (15% or more) despite its broad measurement frequency range (50 ~ 20kHz) [22, 23]. To address this issue, this study adopted the judgment criterion of the maximum amplitude associated with the resonance frequency, a commonly applied approach in conventional traditional low-frequency vibration experiments [24]. Figure 4 illustrates the program block diagram for this aspect.

The program processes the pickup signal and records the frequency and amplitude data acquired at each processing stage in dedicated arrays. The "Array Max & Min" array function is utilized to identify the index position of the highest amplitude value in the amplitude array. Subsequently, the "Index Array" function is employed to extract the resonance frequency from the frequency array using the corresponding index position.

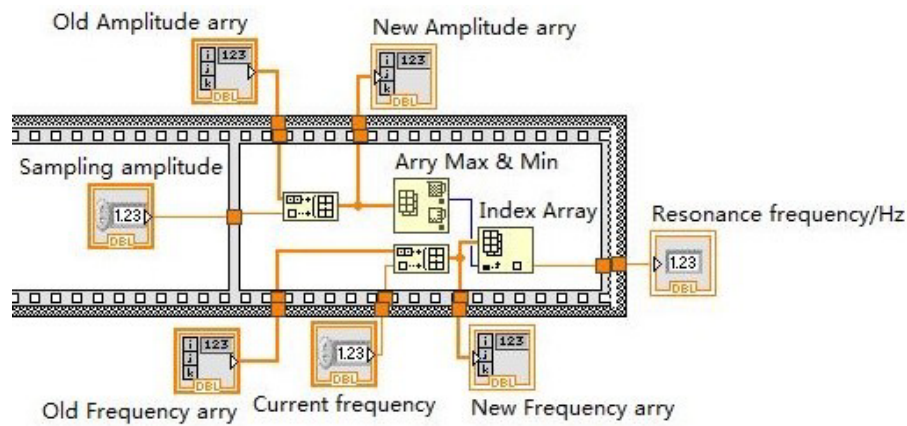

**Figure 4.** Block diagram of the procedure for determining the resonant frequency.

Since the sound card lacks a reference voltage for the analog signal [25], even stable signals may exhibit unstable amplitude values when acquired through the sound card. However, the peak-to-peak value (Vpp) of the signal captured by the sound card still offers a relative assessment of the specimen's vibration amplitude. In the proposed experimental platform, the Vpp of the sound card-acquired signal is used as a basis for data comparison and analysis instead of directly measuring the vibration amplitude of the specimen.

During forced vibration experiments, the specimen must undergo intense oscillations over a specific signal period before reaching a steady-state vibration state [26]. Comparative analysis and assessment of steady-state amplitudes are commonly conducted in low-frequency vibration experiments. To accurately measure the Vpp of the pickup signal, data from the oscillation phase were excluded during fixed frequency measurements, and the effective signal acquisition value was derived from the Vpp of the steady-state vibration. In fixed frequency measurements, the number of signal loop outputs and acquisitions was set to 200. Through multiple experiments, the average Vpp of the last 50 signal acquisition loops was calculated to derive the effective acquisition value. The program block diagram for this component is illustrated in Figure 5.

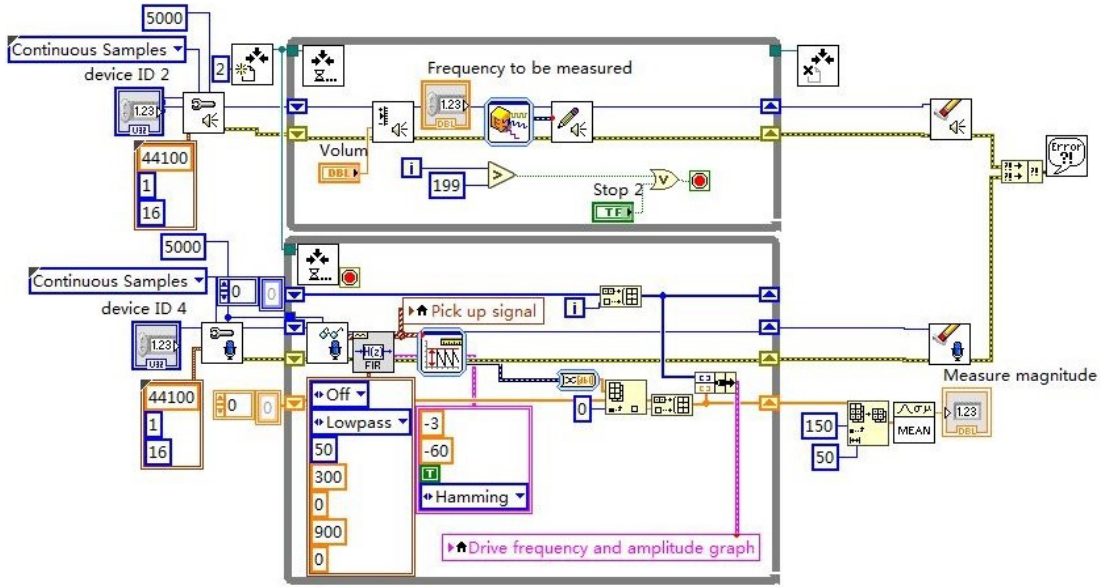

**Figure 5**. Block diagram of the fixed frequency measurement procedure.

To validate the above-mentioned discussion, a low-frequency excitation signal (up to 1200 Hz) was generated using the sound card, and the resulting signal was recorded during specimen excitation through the program. As an example, the waveform of the pickup signal at 257 Hz was analyzed using the "Read Sound Input" function, as depicted in Figure 6(a). The observed Vpp (analog voltage value) was measured at 0.134 V. Figure 6(b) shows the relationship curve between the Vpp of the pickup signal and the acquisition loop number. Experimental investigation revealed that the stability of the pickup signal's Vpp was achieved after 120 acquisition loops (equivalent to a total duration of 2.4 seconds). This outcome serves as evidence supporting the high level of reasonability in the design of the fixed frequency measurement program.

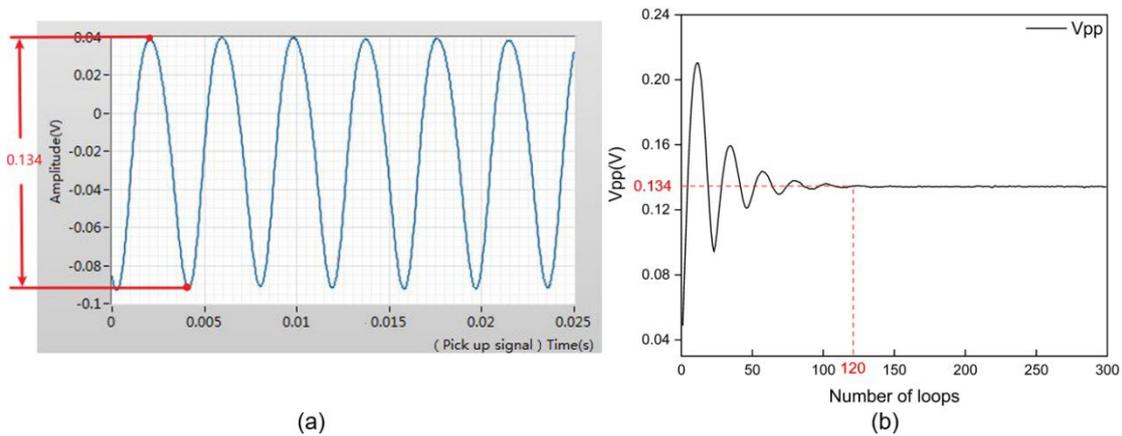

**Figure 6**. Signal reception test of the sound card (a) Pick-up signal waveform (b) Pick-up signal Vpp vs. number of loops.

## 3.5. Dynamic Sweep Program Design and Measurement Using the Three-Point Approximation Method

The dynamic sweep method is widely used in engineering applications [18, 21]. This method enhances measurement efficiency by automatically adjusting the output signal frequency through computer program control. Therefore, the dynamic sweep measurement method was incorporated into low-frequency vibration experiments to acquire an accurate amplitude-frequency relationship chart of the experimental specimen. The platform offered step sizes of 1 Hz, 0.1 Hz, and 0.01 Hz for conducting dynamic sweep measurements.

 During the experiment, it was observed that the feedback pickup signal from the specimen to the computer exhibited significant oscillations when a step size of 1 Hz was used for sweep frequency. To attain steady-state vibration, 150 acquisition loops (equivalent to a total duration of 3 seconds) were required. Conversely, with a medium step size of 0.1 Hz and a small step size of 0.01 Hz for sweep frequency, the signal oscillation was relatively mild, and steady-state vibration was achieved after 100 acquisition loops (2 seconds in total) and 120 acquisition loops (2.4 seconds in total), respectively. Figure 7 shows the dynamic sweep pickup signal observation diagram. The figure demonstrates that the Vpp jitter of the signal during steady-state vibration is more pronounced when employing the large step size compared to the medium and small step sizes.

 Hence, this paper presents a targeted algorithm for determining Vpp (peak-to-peak voltage) in dynamic sweep measurements for different step sizes. For a step size of 1 Hz, the output signals are looped and acquired 200 times at each frequency. The average Vpp of the last 50 acquisition loops is considered the effective acquisition value. In the case of step sizes 0.1 Hz and 0.01 Hz, the output signals are looped and acquired 100 times and 120 times, respectively. The effective acquisition value is determined by extracting the Vpp from the acquisition loop at the 100th and 120th cycles, respectively. To perform this process, LabVIEW's button control and event structure empower users to effortlessly regulate the loop count, step size, and initial cycle number for program sampling at each frequency signal output.

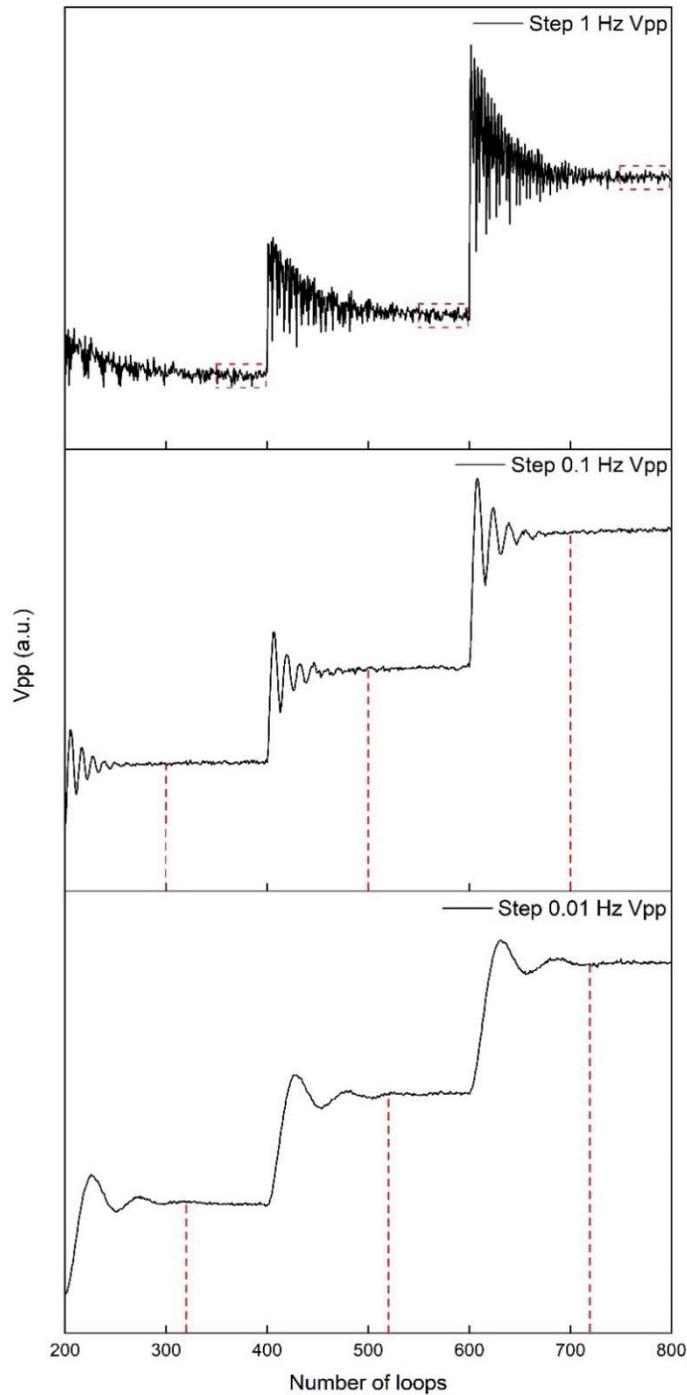

**Figure 7**. The relationship between Vpp and the number of loops during dynamic frequency scanning with different step sizes.

In the experiments focused on forced vibration and resonance, the students investigate a vibrating system centered around a tuning fork as the subject of study. An electromagnet coil generates an electromagnetic force acting as the excitation force, while a piezoelectric transducer measures the vibration amplitude of the forced vibration system with respect to the driving force frequency. This enables the exploration of phenomena and laws of forced vibration and resonance. To determine the quality factor of the tuning fork in this experiment, the students are required to plot a sharpness curve illustrating the relationship between the amplitude of the tuning fork and the resonance frequency. The conventional experimental approach involves measuring the amplitude-frequency relationship across a broad spectrum and visually identifying the frequency range exhibiting significant amplitude fluctuations for precise measurements. However, the conventional method is time-consuming, labor-intensive, and often lacks an adequate number of data points, resulting in an insufficiently smooth sharpness curve. To address these limitations, this study proposes an enhanced dynamic sweep method called the three-point approximation method. The proposed method is constructed based on the previously designed dynamic sweep function. This method significantly improves measurement efficiency, captures more abundant and accurate data points, and consequently yields a sharper and more precise curve.

Specifically, the preliminary sharpness curve is obtained by conducting frequency sweeps with a large step size (1 Hz) to determine the approximate range of the resonance frequency. Subsequently, within this range, frequency sweeps are executed with a medium step size (0.1 Hz) to refine the data points of the sharpness curve, leading to more defined frequency peaks. Finally, near the frequency peak, frequency sweeps are performed with a small step size (0.01 Hz) to achieve a sharper curve with higher precision. The data collected using these different step sizes are consolidated into a single measurement result by utilizing the "delete array elements" and "array insert" controls.

Using the proposed method, the sharpness curve illustrating the relationship between tuning fork Vpp and frequency was obtained within the range of 254 ~ 256 Hz as shown in Figure 8(a). To enhance the visual representation of the measurement data, Origin software was employed to superimpose line graphs depicting the results of three sweep frequency measurements with different step sizes. These line graphs are presented in Figure 8(b), with each step size measurement depicted in a distinct color. The analysis outcomes are highly satisfactory.

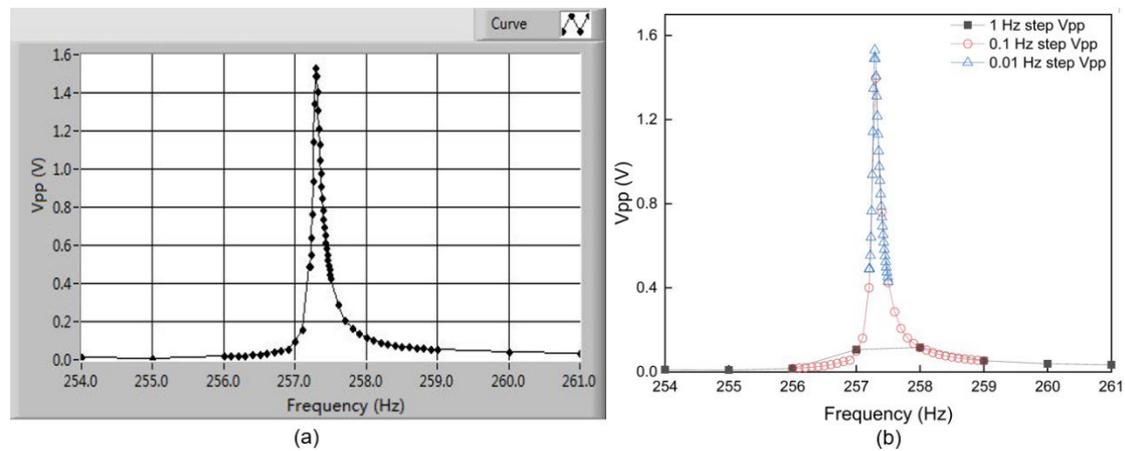

**Figure 8.** Measurement results of the three-point approximation method (a) Relationship between Vpp and frequency as derived from the low-frequency vibration experimental platform program (b) Dot plot illustrated with measurement results from the low-frequency vibration experimental platform program.

## 4. Platform Application Verification

Experiments investigating forced vibration, resonance of tuning forks, and the dynamic measurement of Young's modulus have been successfully conducted using our experimental platform. The experimental results convincingly illustrate the laws under investigation. For a comprehensive overview of the experiments, please consult Supplementary 2.

# 5. Conclusion

This paper presents a detailed exposition of a university-level physics experimental platform designed for studying low-frequency vibration, leveraging virtual instrument technology. The platform streamlines the experimental setup by using only a computer, internal sound card, and essential apparatus, providing an innovative and low-cost solution for low-frequency vibration experiments in educational settings. Comprehensive testing of the LabVIEW-controlled sound card's performance synchronization has convincingly demonstrated that the proposed virtual instrument platform successfully fulfills all the requirements of university-level physics experiments. Expanding on this foundation, various functionalities including fixed frequency measurements, dynamic frequency sweeping, three-point approximation method measurements, and data storage capabilities are developed and incorporated. This platform has been successfully used to perform forced vibration and resonance experiments, as well as dynamic measurements of Young's modulus.

The low-frequency vibration experimental platform developed in this study offers notable advantages in data measurement, data processing, and experiment automation compared with the traditional experimental platforms. The application of this platform in experimental teaching reduces students' manual recording of data, provides students with more efficient experimental methods, and significantly improves the quality of experiments. Moreover, this platform achieves a practical balance between cost and operational performance, offering an alternative avenue for educational applications.

## Acknowledgments

This work was financially supported by Innovation and Entrepreneurship Training Programme for College Students in Jiangsu Province (Grant No.202110291032Z).

## References

1. Kraftmakher Y. Digital storage oscilloscopes in the undergraduate laboratory. *European Journal of Physics* 2012; 33: 1565-1577.

2. Li C, Cao J and Zhang X. Water Temperature Measurement Method Based on Square-Wave Signal Generator. *In: 15th Global Congress on Manufacturing and Management (GCMM 2021) ,Liverpool,UK,25 November -27th November 2020*. J. Phys.: Conf. Ser. 2198 012022.

3. Wu Y, Yang H, Zhang J, et al. Improvement and Practice of Oscilloscope Experiment for Medical Golleges. *Chinese Journal of Medical Physics* 2014; 31: 4967-4970.

4. Williams H. Measuring Young's modulus with a tensile tester. *Physics Education* 2022; 57.

5. Wu J, Yao Y, Chang X, et al. Investigation on the forced vibration characteristics of non-contact elastic swing based on RLC resonant circuit. *College Physics* 2015; 34: 46-49.

6. Perov P, Johnson W and Perova-Mello N. The physics of guitar string vibrations. *American Journal of Physics* 2016; 84: 38-43.

7. Reyes‐Aviles F and Aviles‐Cruz C. Handheld augmented reality system for resistive electric circuits understanding for undergraduate students. *Computer Applications in Engineering Education* 2018; 26: 602-616.

8. Torre Ldl, Jhonnatan S, Dormido S, et al. Two web-based laboratories of the FisL@bs

network: Hooke's and Snell's laws. *European Journal of Physics* 2011; 32: 571-584.

9. Moriarty PJ, Gallagher BL, Mellor CJ, et al. Graphical computing in the undergraduate laboratory: Teaching and interfacing with LabVIEW. *American Journal of Physics* 2003; 71: 1062-1074.

10. Sinlapanuntakul J, Kijamnajsuk P, Jetjamnong C, et al. Computer soundcard as an AC signal generator and oscilloscope for the physics laboratory. *In:Proceedings of the 5th International Conference for Science Educators and Teachers (ISET) 2017,6–8 June 2017,AIP Conf Proc 1923, 030043 (2018).*

11. Huang N, Su G and Li G. Application of the C Language Programming in Measuring Liquid Viscosity by Ball-Dropping Method. *Physical Experiment of College* 2016; 29: 82-85.

12. Lan Y. The application of VB language in the experimental data processing of measuring the moment of inertia of objects by torsion pendulum method. *Science & Technology Information* 2017; 15: 233-234.

13. Liang L and Xie X. Design of Hall effect test system based on LabVIEW. *Laboratory Science* 2018; 21: 26-29.

14. Wang J, Huang L, Wang L, et al. Based on LabVIEW's Malus'Law Experiment. *Physical Experiment of College* 2011; 24: 66-69.

15. Ren W, Jin N and Lei O. Phase Space Graph Convolutional Network for Chaotic Time Series Learning. *IEEE Transactions on Industrial Informatics* 2024: 1-9.

16. Ren W, Jin N and Wang T. An Interdigital Conductance Sensor for Measuring Liquid Film Thickness in Inclined Gas–Liquid Two-Phase Flow. *IEEE Transactions on Instrumentation and Measurement* 2024; 73: 1-9.

17. Ren W and Jin Z. Phase space visibility graph. *Chaos, Solitons & Fractals* 2023; 176.

18. Chen C, Jin H, Feng Y, et al. Data Acquisition Card and Virtual Oscilloscope System. *Instrument Technique and Sensor* 2012; 49: 67-69+72.

19. Li X and Fan X. A Data Acquisition System of Virtual Instrument Based on Sound Card and Java. *Journal of Henan Normal University(Natural Science Edition)* 2006; 47: 174 -176.

20. Zhu D. Vocal Frequency Spectrum Analytical Instrument Based on the Technology of Virtual Instrument. *Journal of Hefei University(Comprehensive ED)* 2005; 22: 91-93.

21. Song M. *LabVIEW Programming in Detail*. Publishing House of Electronics Industry, 2017,p.293-294.

22. Cao H. *Research on Audio signal Filtering and Recognition Technology based on time-frequency domain analysis*. Master Thesis, Guangzhou University, China, 2016.

23. Sun J, Gu X and Qi F. Research Rectangular Reed Resonance Frequency. *Metrology & Measurement Technique* 2015; 42: 41-43.

24. Ding S, Chen J and Zhu X. Designing innovative experiments to cultivate students' core literacy -- Taking forced vibration resonance teaching as an example. *Physics Teaching* 2022; 44: 23-26.

25. Wang M. *The Design of a Virtual Oscillograph Based on Sound Card*. Master Thesis, Taiyuan University of Technology, China, 2006.

26. He X, Ren Y, Mao W, et al. Graphic Method for Quantitative Study of Forced Vibration Experiment. *Physical Experiment of College* 2002; 35: 112-116.

# Supplementary file 1. Experimental Testing of Sound Card Signal Transmission and Reception Performance

## 1. Feasibility Analysis of Sound Card Signal Transmission and Reception

  University physics experiments include several low-frequency vibrations experiments, such as forced vibration and resonance experiments (frequency range: 0 ~ 800 Hz) [1], dynamic method measurement of Young's modulus experiment (frequency range: 500 ~ 1000 Hz) [2], and string vibration experiment (frequency range: 50 ~ 200 Hz) [3]. The excitation and measurement signals required for these experiments are all sine waves, with frequencies limited within the range of 0 ~ 1200 Hz. However, the oscilloscopes and signal generators currently used in laboratories surpass the requirements of these experiments in terms of input and output capabilities, leading to increased costs resulting from their high prices. Considering the standardized integration of sound cards within computer systems, their input and output signal frequency range (0 ~ 44 kHz) adequately satisfies the prerequisites of low-frequency vibration experiments. [4] Moreover, they are also comparatively cost-effective. Therefore, the potential substitution of expensive signal generators and oscilloscopes with affordable sound cards for conducting low-frequency vibration experiments holds the promise of ensuring experiment effectiveness while simultaneously curbing overall expenses. Based on relevant studies, this paper investigates the use of sound cards in low-frequency vibration experiments and analyzes their feasibility in the field of university physics.

  The signal input and output performance of the sound card was assessed and validated through experiments using the RIGOL DG4062 Digital Signal Generator (with a signal frequency range of 0-60 MHz) and the RIGOL DS1102 Digital Oscilloscope (with a signal acquisition frequency range of 0-50 MHz) manufactured by RIGOL TECHNOLOGIES, Co. LTD.

  Typically, the playback channel of a computer sound card is utilized for outputting multimedia audio signals. In coordination with the computer's microprocessor and sound effect software, instructions and data are transmitted to the sound card's sound processing chip. This communication effectively manages various components, including digital-to-analog converters, enabling the generation of the desired sound signals [5]. To achieve a precise substitute for the signal generator, the waveform outputted by the sound card, which is controlled by the program, must exhibit both accuracy and stability. In this study, the sine wave output of the sound card across various frequencies ranging from 0 to 1200 Hz was observed using the CH1 channel of a digital oscilloscope. Additionally, the waveform output of a signal generator operating at the same frequencies was observed using the CH2 channel of the same oscilloscope for comparative purposes. Figure 1 portrays the waveforms recorded by the oscilloscope at 100 Hz and 1000 Hz frequencies. Overall, the findings indicate that the sine wave signal produced by the sound card remains largely consistent with the output of the signal generator within this frequency range.

  The power of the audio signal directly outputted by the playback channel of a sound card typically falls within the range of 20-100 mW (as per international standards). However, this power level is inadequate for inducing vibration in the experimental specimen. Consequently, the experimental platform employed the TDA2030A audio amplifier module to magnify the power of the output signal generated by the sound card. Again, the signal at the rear end of the TDA2030A audio amplifier module was monitored using the CH1 channel of a digital oscilloscope (Figure 1). The figure reveals that the amplitude of the sine wave signal is increased after power amplification. Notably, the phase and frequency remain unaltered, and no significant noise is observed.

Simultaneously, in the experiment, the signal following power amplification effectively induced noticeable vibrations in the specimens throughout the forced vibration, resonance and dynamic Young's modulus measurement experiments.

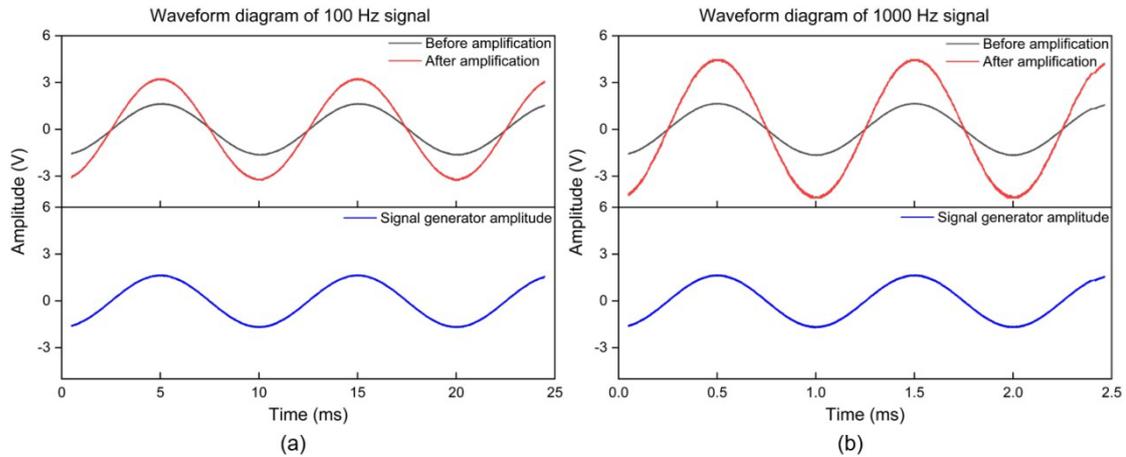

**Figure 1**. Comparison of output signals from pre- and post-amplification of the sound card with signal generator output at frequencies (a) 100 Hz and (b) 1000 Hz.

The recording channel of a sound card is responsible for receiving analog signals from audio devices, such as microphones. After undergoing filtering and amplification, the received analog signal is routed to an ADC that samples the analog signal at a fixed time interval determined by the sampling rate clock frequency. This process converts the analog signal into a digital format recognized by the computer [5]. The following experiments were conducted to evaluate the feasibility of replacing a digital oscilloscope with the sound card's recording channel and meeting the signal acquisition needs in low-frequency vibration experiments. In this study, a digital signal generator was used as the signal source to produce standardized signals of varying frequencies (sine waves with an amplitude of 1.2 V) within the range of 0 to 1200 Hz. A program designed by LabVIEW was implemented on the computer to control the sound card's recording channel, enabling real-time display and measurement of the standardized signals. Figure 2 illustrates the real-time results obtained from capturing signals at 100 Hz and 1000 Hz frequencies. The measurement results indicate that the "single-frequency measurement" function accurately corresponds to the frequency outputted by the signal generator. It is worth noting that the recording channel of a computer sound card converts received electrical signals into analog signals with voltage values confined within a ±1 V range [6]. Consequently, the signal amplitude captured by the sound card approximately reaches 0.86 V, slightly deviating from the output signal amplitude of the signal generator (1.2 V). However, this slight deviation has no impact on the experiments conducted in this study. In conclusion, the comparative analysis of signal frequency characteristics validates that the signal acquisition capabilities of the sound card satisfactorily fulfill the demands of low-frequency physics experiments within the range of 0 to 1200 Hz.

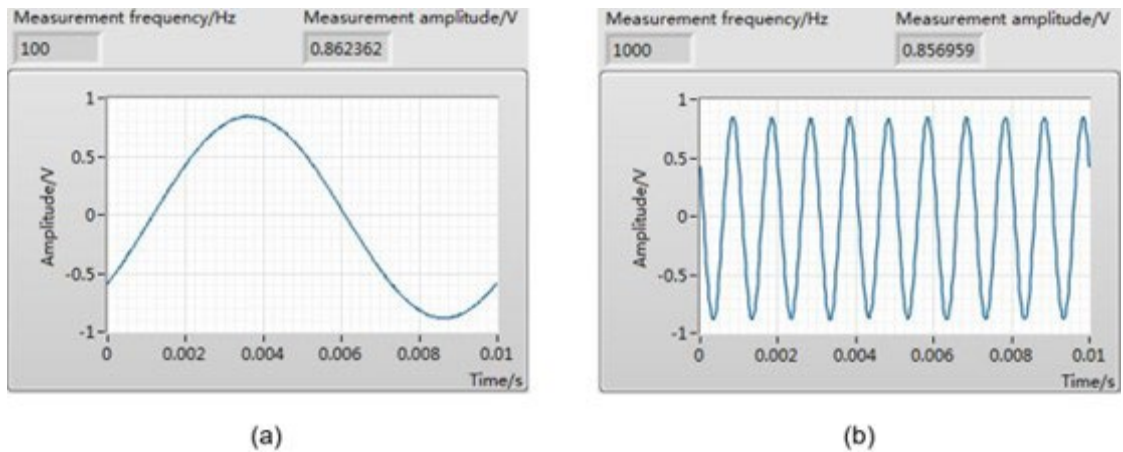

**Figure 2**. Signal acquisition test results of the sound card at frequencies (a) 100 Hz and (b) 1000 Hz.

## 2. The response speed of the system

The response speed of the system serves as a pivotal parameter in instrument measurement. In our initial manuscript, we scrutinized and evaluated the time response between the input and output of the sound card signal, which is 0.22 ms (as referenced in page 6, lines 17 - 19 of the main document). Unfortunately, we overlooked the overall system response time. We are grateful to the reviewer for pointing out this oversight. Consequently, in the revised version of Supplementary File 1, we have incorporated a thorough analysis of the system response of this platform in the second section:

We developed an experiment program using LabVIEW graphical programming language to measure the time($\Delta t$) it takes for an analog signal emitted by the sound card to traverse the sample and subsequently received by the same sound card. As Figure 4 demonstrates, the measured $\Delta t$ was approximately 0.0125 s. This time is considered the response time in this system. Additionally, given that the platform necessitates a minimum of 2 - 3 s to acquire each data point (see page 9, lines 2 - 11 of the main document), which significantly exceeds $\Delta t$, we believe that this system delay does not significantly impact the measurement results.

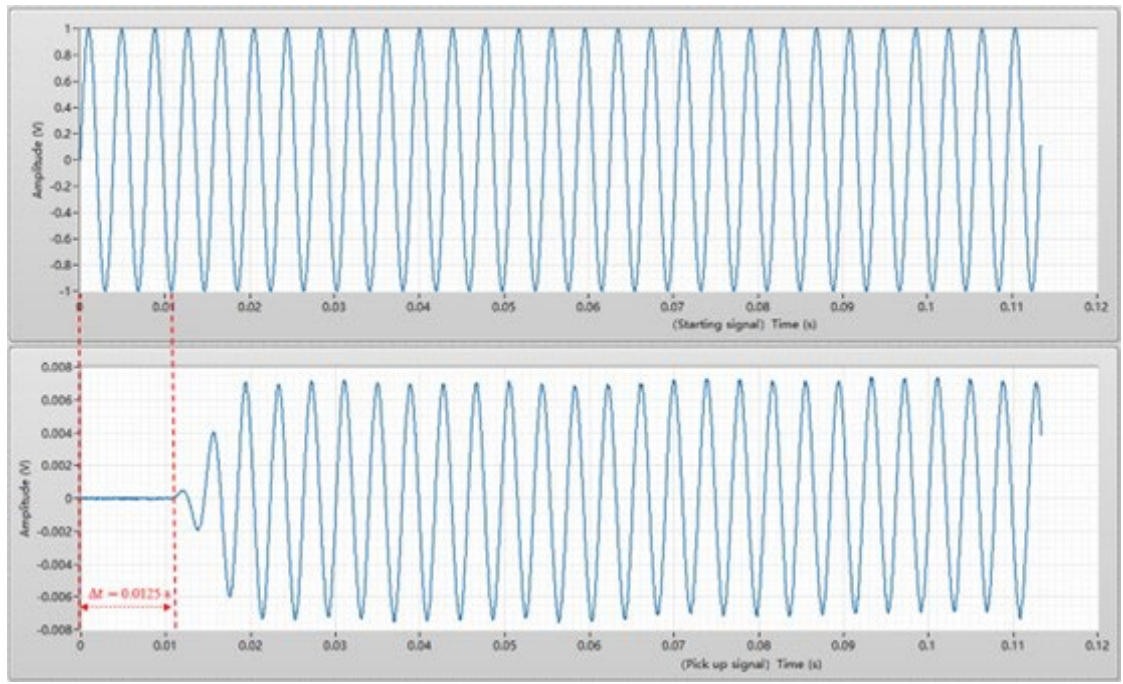

**Figure 3.** Response Speed Analysis of System Test Diagram

## References

1. Ding S, Chen J and Zhu X. Designing innovative experiments to cultivate students' core literacy -- Taking forced vibration resonance teaching as an example. *Physics Teaching* 2022; 44: 23-26.
2. Huang Q, Qian Z, Peng Y, et al. A New Method for Measuring Young's Modulus of Solid Materials by Dynamic Method. *Physical Experiment of College* 2018; 31: 25-28.
3. Wang X, Fu Y, Zhu Y, et al. Study on the Influence of Chord Length on String Vibration Experiment. *Physical Experiment of College* 2022; 35: 39-41.
4. Yang Z. Design of Sine Wave Audio signal Generator. *Audio Engineering* 2004; 28: 35-36.
5. Rao Z. Sound card working principle. *Electronic Test* 2000; 07: 156-157+160.
6. Chen C, Jin H, Feng Y, et al. Data Acquisition Card and Virtual Oscilloscope System. *Instrument Technique and Sensor* 2012; 49: 67-69+72.

# Supplementary file 2. Verification of platform application through experiments investigating forced vibration, resonance of tuning forks, and dynamic measurement of Young's modulus

## 1. Experimental data and results

In Section 3.5 of the main article, the three-point approximation method within the measurement function of the low-frequency vibration experimental platform was applied to evaluate the sharpness curve depicting the variation of the tuning fork's forced vibration and resonance frequency in relation to Vpp (Figure 8(a) in the main article). Moreover, the automated analysis feature of the platform program was utilized to acquire the resonance frequency($f_0$) and the half-power frequencies ($f_1$ and $f_2$), as illustrated in Figure 1. Subsequently, these values were employed to calculate the quality factor (Q)(As shown in Equation (1)). Notably, the platform has the capability to automate this calculation and provide the corresponding result.

$$Q = \frac{f_0}{f_2-f_1} = \frac{257.29}{257.34-257.26} = 3216.125 \tag{1}$$

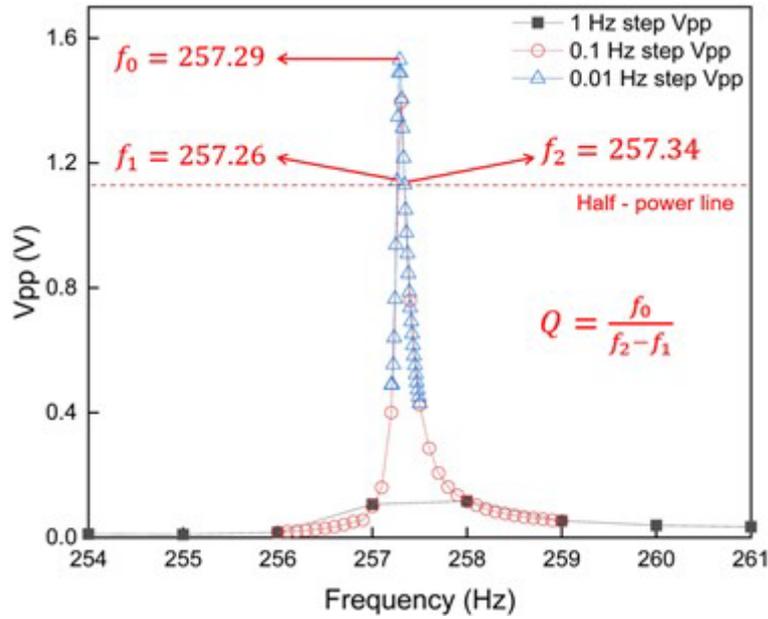

**Figure 1.** Sharpness curve of tuning fork measured by the low-frequency vibration experimental platform program.

Next, we used the low-frequency vibration platform to acquire and present a series of data point diagrams illustrating the relationship between the vibration signal amplitude of the tuning fork and its frequency response, when a 5 g weight was loaded at distances of 2, 3, 4, 5, 6, 7, and 8 cm from the tuning fork's end, as shown in Figure 2.

As previously mentioned in the original manuscript, the resonant frequency is determined by comparing the received signal data, specifically identifying the frequency corresponding to the maximum value (see lines 3 - 6 on page 7 of the Main Document). Our platform facilitates the automatic processing of image data points, accurately positioning the resonant frequencies on the user's interface. Ultimately, the platform was able to fit these data to derive an experimental rule, depicted in Figure 1(h). The formula derived from the platform's fitting of the data points is presented in formula (2). The experimental results indicate that the position of the tuning fork's loading weight and the resonance frequency adhere to a quadratic curve relationship. The experimental law proposed by the platform aligns with the findings of other research groups. [1]

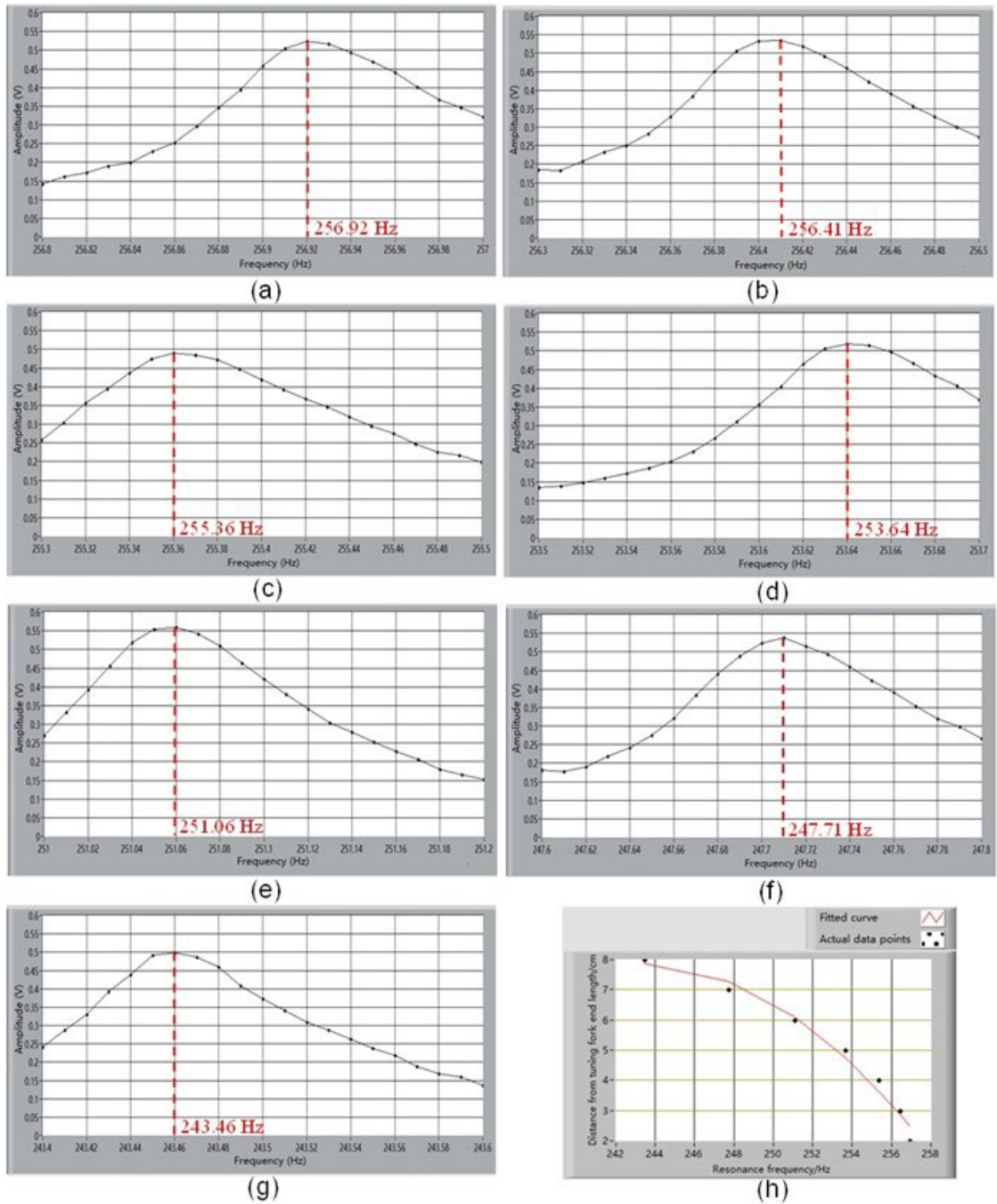

**Figure 2.** Peak value vs. frequency of tuning fork vibration signal obtained from the low-frequency vibration experimental platform with 5g mass weight loaded at distances of (a) 2 cm, (b) 3 cm, (c) 4 cm, (d) 5 cm, (e) 6 cm, (f) 7 cm, and (g) 8 cm from the fork's end.

$$y = -0.0287165x^3 + 13.9728x^2 - 1691.83 \qquad (2)$$

Then, we apply the platform to investigate the relationship between the loading quality of the tuning fork and its resonance frequency. Figure 3 shows the fitting diagram of the experimental data provided by the platform, along with the fitting curve formula (3). The results indicate that the relationship fitted by the platform aligns with a linear pattern. [1]

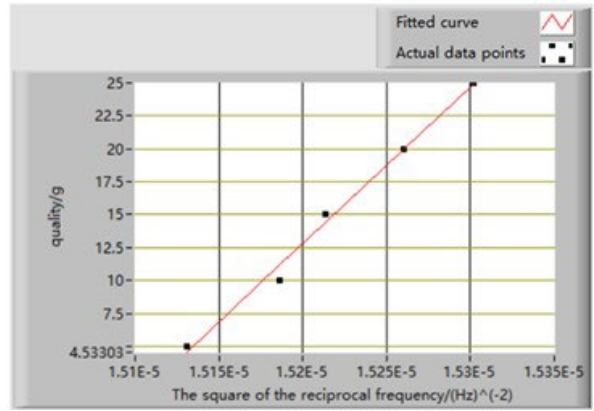

**Figure 3.** The fitting diagram of the relationship between the mass of the loaded weight and the resonant frequency of the tuning fork at a distance of 2cm from the end of the tuning fork shown by the platform

$$y = 1.28799 \times 10^8 x - 1944.11 \tag{3}$$

The aforementioned experimental results indicate that the impact of load position and mass on the resonant frequency of a tuning fork can be harnessed for measurement applications.

The dynamic method is a widely utilized experimental technique for measuring Young's modulus, which determines the elastic properties of materials. This technique entails applying periodic external forces to the material and measuring its vibrational response to determine Young's modulus. In Young's modulus experiment conducted using the dynamic method measurement, the dynamic sweep function of the platform was used to effectively illustrate the relationship between the resonance frequency of the test rod and the distance between the excitation point and the pickup point at the rod's end. Iron, aluminum, and glass rods were selected as test rods for the experiment. The experimental results are depicted in Figures 4(a), 2(b), and 2(c). It can be seen from the figure that the proposed platform's extrapolation yielded a resonance frequency ($f_r$) of 4.48 cm for the test rod, which aligns with the frequency measured by conventional experimental instruments.

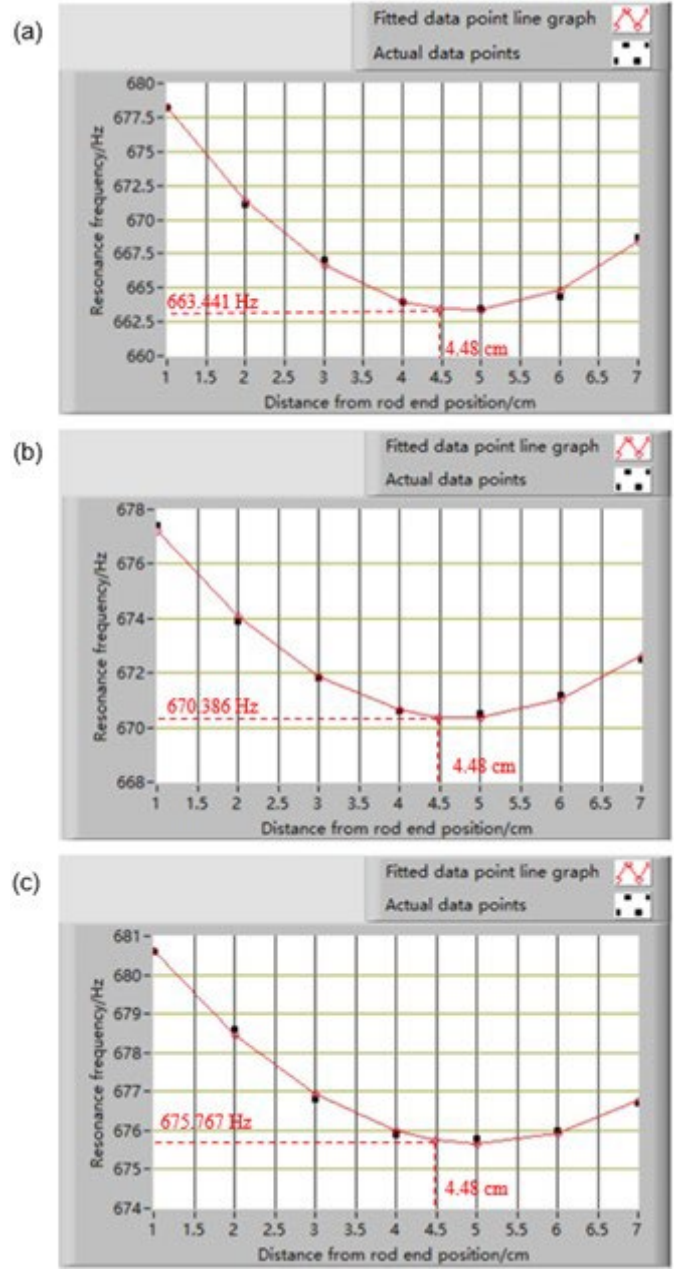

**Figure 4.** Experimental results of the low-frequency vibration experimental platform applications for the dynamic measurement of Young's modulus, including: (a) Young's modulus of iron rod measured by extrapolation method (b) Young's modulus of aluminum rod measured by extrapolation method (c) Young's modulus of glass rod measured by extrapolation method.

The resonant frequencies of the three materials, measured using the platform, were substituted into equation (3) to calculate their respective Young's moduli. Notably, the calculation process is facilitated through the automated capabilities of the platform.

$$E = 1.6067 \times \frac{ml^3 f_r^2}{d^4} \times T_1 \tag{3}$$

In this equation, m, l, and d represent the mass, length, and diameter of the test rod, respectively, while $T_1$ is the correction factor, whose value is dependent on the ratio of d/l. For all three materials, a value of 1.005 was adopted for $T_1$. [2]

The physical parameters of the sample rods for the three materials, as well as the calculated Young's modulus using the appropriate formula, are presented in Table 1. The measured Young's moduli for the materials are compared against their respective reference values presented in Table 1. The relative error of the measured Young's moduli was within 3%, whereas the traditional method yielded a relative error of 8%.[3] This demonstrates the high accuracy of our platform in data processing.

Table 1. Comparison of experimental results of Young's modulus with reference values.

| Material | Mass (g) | Length (mm) | Diameter (mm) | $f_r$ (Hz) | $E_e$ ($10^{10}$N/m$^2$) | $E_r$ ($10^{10}$N/m$^2$) | RE |
| --- | --- | --- | --- | --- | --- | --- | --- |
| Steel | 44.4 | 200 | 5.961 | 663.441 | 20.01 | 20.6 | 2.8% |
| Aluminum | 15.3 | 200 | 5.972 | 670.386 | 6.99 | 7 | 0.1% |
| Copper | 11.744 | 200 | 5.976 | 675.767 | 5.43 | 5.29 | 2.6% |

$f_r$: resonance frequency. $E_e$: experimental value of young's modulus. $E_r$: Reference value of Young's modulus. RE: relative error.

The experimental verification demonstrates the ease of operation and convenient usability of the proposed low-frequency vibration experimental platform. This platform offers various functionalities including fixed frequency measurement, dynamic sweep measurement, real-time graphical display of measurement data, storage and analysis of results and presentation of experimental findings. It seamlessly integrates computer technology, software, and instruments, utilizing virtual instruments to significantly assist measurement personnel and enhance their capabilities.

## 2. Systematic error analysis

It is well known that measurement systems inherently possess systematic errors during the measurement process. Obviously, our low-frequency vibration experiment platform is also subject to systematic errors.

In our low-frequency experiment conducted on the platform, various methodologies were used to measure diverse physical quantities. Specifically, the mass, diameter, and length of the sample were measured using a balance, spiral micrometer, and vernier caliper, respectively. For example, the diameter of the sample rod was accurately measured using a spiral micrometer, achieving a precision of 0.01 mm.

In the experiment, the sound card's (Realtek ALC662 audio chip) sampling rate was set to 96 kHz with a sampling depth of 24 bits. [4] This configuration ensures that the resolution of the output signal in terms of frequency and amplitude significantly exceeds the requirements of the experiment. Furthermore, through manual programming by LabVIEW, the output and input signals of the sound card are precisely controlled, resulting in a frequency resolution of up to 0.001 Hz. The aforementioned measuring devices, as components of the low-frequency vibration experimental platform, are conformed to introduce systematic errors.

In university physics experiments, the graphing method is a prevalent and efficient approach for data processing. For instance, in the tuning fork forced vibration and resonance experiment, graphing experimental data aids students in understanding experimental data and phenomena. In the dynamic measurement of Young's modulus, the resonance frequency of the sample rod is determined through extrapolation methods. However, the graphing method is inherently prone to introducing errors, and evaluating the magnitude of these errors poses a more intricate challenge. In experiments employing the graph method, the relative deviation, typically assessed by comparing the experimental result with the accepted value, is considered satisfactory if it falls below a predetermined threshold.[5] Table 1 demonstrates that our measurements exhibit a relative error of less than 3%, significantly outperforming traditional experimental methods in terms of accuracy. [3] Additionally, our experimental results indicate that the system error of our low-frequency vibration experimental platform is significantly lower than that of traditional experimental instrument systems.

**References**

1. Guo Y. Experimental research on tuning fork's forced vibration. *Laboratory Science* 2020; 23: 40-43.
2. Gu D and Wu G. *College Physics Experiment*. Shanghai Jiao Tong University Press, 2018，p.153-154.
3. Congcong F, Huayang L and Mingxi C. Improvements of Measuring the Young's Modulus by Dynamic Method and Its Research on Teaching. *Physical Experiment of College* 2016; 29: 51-59.
4. Yang Z. Design of Sine Wave Audio signal Generator. *Audio Engineering* 2004; 28: 35-36.
5. Chen X. Plot and error. *Physics Experimentation* 1994; 14: 265-266.